\documentstyle[11pt]{article}
\begin{document}
\def\theequation{\arabic{section}.\arabic{equation}}
\newcommand{\be}{\begin{equation}}
\newcommand{\ee}{\end{equation}}
\begin{titlepage}
\title{Laser interferometric detectors \\ of gravitational waves}
\author{F. I. Cooperstock      \\ \\ and \\ \\
V. Faraoni \\ \\{\small \it Department of Physics and Astronomy, University
of Victoria} \\
{\small \it P.O. Box 3055, Victoria, B.C. V8W 3P6 (Canada)}}
\date{}
\maketitle   \thispagestyle{empty}  \vspace*{1truecm}
\begin{abstract}
A laser interferometric detector of gravitational waves is studied and
a complete solution (to first order in the metric perturbation) of the
coupled Einstein-Maxwell equations with appropriate boundary conditions for
the light beams is determined. The phase shift, the light deflection
and the rotation of the polarization axis induced by gravitational waves
are computed. The results are compared with previous literature, and
are shown to hold also for detectors which are large in comparison with the
gravitational wavelength.
\end{abstract}  \vspace*{1truecm}
\begin{center}
To appear in {\em Classical and Quantum Gravity}
\end{center}  \end{titlepage}   \clearpage
\section{Introduction}
The detection of gravitational waves is an important goal of
contemporary physics and in recent years, research in the field has
shifted towards laser interferometric detectors
(\cite{Thorne}-\cite{Giazotto} and references therein). Large laser
interferometers
are presently under construction and in the planning stages in various
countries. Remarkable technological progress has been made in the effort to
detect the weak gravitational waves that are believed to be emitted from
extragalactic sources.

The analysis of a laser interferometric detector is usually carried
out in Fermi normal coordinates (FNC), which are associated with a
freely falling observer carrying the detector \cite{MTW,Thorne}. In order
to introduce these coordinates, one must restrict oneself to
considering detectors with size $ a $ much smaller than the reduced
wavelength $\lambda/2\pi $ of the gravitational waves to be detected.
This condition is satisfied only marginally in some cases
\cite{Thorne}, and as larger interferometers are being planned, it
will be satisfied even less and ultimately, not at all in certain
frequency bands. Most detectors are
conceived to operate at gravitational wave
frequencies $ \nu_{g} \leq 10$ KHz, with maximum sensitivity
up to $1$ KHz. The Italian-French project VIRGO plans the
construction of an interferometer with arm size $ a \sim 3 $ Km, and
maximum sensitivity around $ 100$ Hz. The detector in its final stage
is expected to operate in the range from $10$ Hz to $1$ KHz.
For $ \nu_{g} \sim 1$ KHz, $ 2\pi a/\lambda_g \simeq 0.06 $,
and in the extreme case $ \nu_g \sim 10$ KHz, $2\pi a /\lambda_g
\simeq 0.6 $. Beam detectors in space for the detection of low frequency
gravitational waves ($ \nu_g \sim 10^{-1} $ Hz) have also
been proposed (\cite{Weiss1,Weiss2}; \cite{Thorne} and references therein),
although the necessary technology is not available at present; for these
systems, $ a\sim 10^{6} $ Km and $ \nu_g \sim 10^{-1}$ Hz give $ 2\pi
a/\lambda_g \sim 2 $. Therefore, it appears convenient to generalize
known results to the case $ a \geq \lambda_g /2\pi $. Moreover, since energy
localization and transfer in general relativity are not
in as nearly a well-established state as they are in the rest of physics,
it is particularly important that the theoretical predictions for the
effects of gravitational waves in their interaction with proposed
detectors be established with great care. The advantage of laser
interferometry as a detection mechanism is that it requires no reference to an
energy transfer mechanism (or even whether such a transfer actually
occurs \cite{Cooperstock92})  but rather, can be phrased entirely in terms of
phase shifts. In fact, this carries with it the additional advantage of
being expressible in terms of scalar quantities upon which all observers
must agree. In a recent paper, Lobo \cite{Lobo} usefully focussed upon
these aspects.

The formalism for the interference between electromagnetic and gravitational
waves was established many years ago \cite{Cooperstock68}
and the perturbations to the electromagnetic waves in the case of weak
fields were
solved for longitudinal and transverse orientations. Moreover, they were
solved with reference to well-defined boundary conditions. In Lobo's work,
the phase shift is calculated for a wave propagating in one direction and
this is added to the phase shift for a wave travelling in the opposite
direction. However, to correspond to the actual physical configuration of
a laser interferometer, it is necessary to be more specific and analyze
the response to a gravity wave of electromagnetic waves which travel
equal distances along the interferometer arms at right angles to each other
and then reflect at the mirrors which terminate these arms. Only when this
is done can one have confidence in the expected interference which one
seeks as the waves are recombined. In this paper, we use the
electromagnetically gauge-invariant perturbation formalism of
Ref.~\cite{Cooperstock68} in conjunction with the boundary conditions
appropriate to a laser interferometer, to derive the expected phase
shift as the reflected waves recombine with interference. Using this
formalism, we are also able to compute the extent of the deflection of
the laser beams due to the interference with the gravitational waves.
While this is a first order effect in the metric perturbation, it only
induces a second order contribution to the phase shift. The value of
the deflection angle is then compared with the results from previous
analyses made in a different context. In addition, we show that
gravitational waves do not induce any rotation in the polarization
plane of electromagnetic radiation.

The plan of the paper is as follows: in Sec.~2 an idealized model of
an interferometric detector is described, and the Maxwell equations in
the presence of a weak, monochromatic, optimally polarized
gravitational wave are solved for an example case. In Sec.~3, the solutions
for the actual interferometer are derived from this case by means of
suitable coordinate transformations, and the phase shift in the light
coming from the two interferometer arms is computed. In Sec.~4, we
take into account the deflection of light beams and changes in the
polarization induced by gravitational waves, showing that these
effects are negligible. Finally, the phase shift is compared with previous
literature in Sec.~5.
\section{An idealized beam detector}
We consider a simplified model of a laser interferometric detector,
whose geometry is given in Fig.~1. The detector consists of a Michelson
interferometer, in which three masses are hanging in the plane $ z=0$.
The periods of the suspension pendula are much greater than the
gravitational wave period $ T_{g}$, so the masses can be regarded as
freely falling. A beam splitter at the origin  splits the light from a
laser lamp into two beams of equal intensity which travel along the
$x$ and $y$ axes, and are reflected by mirrors placed at a
distance $ a $ from the origin, and perpendicular to the $x$ and $y$
axes, respectively. The reflected beams are collected at a photodiode
which compares them and detects any differential effect induced by an
impinging gravitational wave in the interferometer arms.

We assume that (see Ref.~\cite{Lobo} for a discussion of these
assumptions)
\begin{itemize}
\item the electromagnetic field in the laser beam can be treated as a
test field;
\item the mirrors at $ x=a$, $y=a$ are perfectly reflecting;
\item the transverse dimension of the light beam is much smaller than
the gravitational wavelength $\lambda_g$;
\item the light frequency is much larger than the gravitational wave
frequency,\setcounter{equation}{0}
\be
\omega >>\omega_g \; .                      \label{2.1}
\ee
\end{itemize}
We consider a monochromatic gravitational wave travelling along the
positive $z$ axis with a single polarization, and with polarization
axes along the ``L'' of the interferometer. The spacetime metric is
given by\footnote{The metric signature is $+2$, and $ c$ denotes the
velocity of light in vacuum. Greek indices run from
0 to 3 and Latin indices run from 1 to 3. We will perform computations to
first order in the metric perturbation $ h_{\mu\nu}$.}
\be
g_{\mu\nu}=\eta_{\mu\nu}+h_{\mu\nu}              \label{2.2}
\ee
in an inertial frame of the Minkowskian background metric $\eta_{\mu\nu} $,
and $ |h_{\mu\nu}|<<1$. For a polarized beam in the TT gauge, the only
nonvanishing $ h_{\mu\nu}$ describing the gravitational wave are
\be
h_{11}=-h_{22}=-A \cos( k_g z-\omega_g t) \; ,        \label{2.3}   \ee
where $ A $ is a (small) constant and $ k_g=\omega_g/c$.
Although we will restrict our attention to this particular direction of
propagation and to this polarization of the gravitational wave, the
calculation can be generalized to arbitrary incidence and polarization,
in the manner of Ref.~\cite{Lobo}.

The unperturbed electromagnetic field in the $x$ arm of the
interferometer has the only nonvanishing components
\begin{eqnarray}
& & E^{(0)}_{y}=-F^{(0)}_{02}=E_{0}\left[ e^{i(kx-\omega
t)}-e^{-i(kx+\omega t-2ka)} \right] \; ,            \label{2.4} \\
& & H^{(0)}_{z}=F^{(0)}_{12}=E_{0}\left[ e^{i(kx-\omega
t)}+e^{-i(kx+\omega t-2ka)} \right] \; ,            \label{2.5}
\end{eqnarray}
where $ E_0=$constant, $ k=\omega/c$, and where only the real components of
the complex fields need to be retained. The unperturbed field in the $y$ arm
is given by
\begin{eqnarray}
& & E^{(0)}_{x}=-F^{(0)}_{01}=-E_{0}\left[ e^{i(ky-\omega
t)}-e^{-i(ky+\omega t-2ka)} \right] \; ,           \label{2.6} \\
& & H^{(0)}_{z}=F^{(0)}_{12}=E_{0}\left[ e^{i(ky-\omega
t)}+e^{-i(ky+\omega t-2ka)} \right] \; .            \label{2.7}
\end{eqnarray}
These fields satisfy the flat space Maxwell equations
\be
{F^{(0) \, \mu\nu}}_{,\nu}=0 \; ,  \label{2.8}  \ee
\be
F^{(0)}_{\mu\nu ,\rho}+F^{(0)}_{\nu\rho ,\mu}+F^{(0)}_{\rho\mu ,
\nu}=0 \; ,                     \label{2.9}  \ee
with perfectly reflecting boundary conditions at the mirrors located
at $(a,0,0)$ and $(0,a,0)$ (see Fig.~1). We decompose the electromagnetic
field tensor as the sum of the unperturbed field and a small perturbation
induced by the gravitational wave
\be
F_{\mu\nu} \equiv F^{(0)}_{\mu\nu}+F^{(1)}_{\mu\nu} \; ,   \label{2.10} \ee
with $ |F^{(1)}_{\mu\nu}|<<|F^{(0)}_{\mu\nu}|$. The field equations for
the perturbations $ F^{(1)}_{\mu\nu}$ in the TT gauge have been obtained in
Ref.~\cite{Cooperstock68}:
\be
F^{(1)}_{\mu\nu ,\rho} \, \eta^{\rho \nu}={h_{\mu}}^{\nu ,\rho}
F^{(0)}_{\nu\rho}+h^{\nu\rho}F^{(0)}_{\mu\nu ,\rho}+O(h^2) \; ,  \label{2.11}
\ee
\be
F^{(1)}_{\mu\nu ,\rho}+F^{(1)}_{\nu\rho ,\mu}+F^{(1)}_{\rho\mu ,
\nu}=0 \; .                 \label{2.12}  \ee
It has been shown in Ref.~\cite{Baronietal} that other expressions in
the literature are incorrect because they do not account properly for the
Lorentz condition\footnote{Note that Eqs.~(\ref{2.11}), (\ref{2.17})
and (\ref{2.20}) are linear both in $ h_{\mu\nu}$ and in
$ F^{(0)}_{\mu\nu}$, but not in these two kinds of variables
simultaneously. As a consequence, one cannot use the complex
representations for both fields at the same time, retaining only their
real parts in the final results. However, one is always allowed to
make use of the identity $\cos z=\left( e^{iz}+e^{-iz}\right) /2$.}. Equations
(\ref{2.11}) and (\ref{2.12}) must be
solved in both arms using perfectly reflecting boundary conditions. We
solve these equations in a special orientation which does not correspond to
an actual interferometer arm  in our geometry depicted in fig.~1. We
then obtain the solution to Eqs.~(\ref{2.11}) and (\ref{2.12}) for
the actual interferometer arms by means
of suitable coordinate transformations. The fictitious system is
composed of an electromagnetic wave propagating along the $z$ axis,
from $z=0$ to $z=a$, where it is reflected by a perfect mirror;
the system is perturbed by a gravitational wave described by
\be
h_{11}=-h_{33}=A\cos( k_g y-\omega_g t) \; .     \label{2.13}  \ee
The unperturbed electromagnetic field is given by
\be
E^{(0)}_{x}=E_{0}\left[ e^{i(kz-\omega
t)}-e^{-i(kz+\omega t-2ka)} \right] \; ,         \label{2.14}  \ee
\be
H^{(0)}_{y}=E_{0}\left[ e^{i(kz-\omega t)}+e^{-i(kz+\omega t-2ka)}
\right] \; .                                    \label{2.15}  \ee
Equations (\ref{2.11}) for the electromagnetic tensor perturbations
are
\begin{eqnarray}
& & F^{(1)}_{01,1}+F^{(1)}_{02,2}+F^{(1)}_{03,3}=0 \; ,  \label{2.16} \\
& & -F^{(1)}_{10,0}+F^{(1)}_{12,2}+F^{(1)}_{13,3}=h_{11,0}F^{(0)}_{01}-
h_{11}F^{(0)}_{13,3} \; , \label{2.17} \\
& & -F^{(1)}_{20,0}+F^{(1)}_{21,1}+F^{(1)}_{23,3}=0 \; , \label{2.18} \\
& & F^{(1)}_{30,0}+F^{(1)}_{31,1}+F^{(1)}_{32,2}=0 \; ,   \label{2.19}
\end{eqnarray}
from which we can easily derive an inhomogeneous wave equation for
$F^{(1)}_{01}$:
\begin{eqnarray}
& & F^{(1)}_{01,00}-F^{(1)}_{01,22}-F^{(1)}_{01,33}=h_{11,00}F^{(0)}_{01}+
h_{11,0}\left( F^{(0)}_{01,0}-F^{(0)}_{13,3} \right)
-h_{11}F^{(0)}_{13,30}= \nonumber \\
& & =\frac{AE_0}{2}\left\{ (K^+)^2 \left[ e^{i(kz+k_g y-\Omega^+ t)}
-e^{-i(kz-k_g y+\Omega^+ t-2ka)} \right] \right. \nonumber \\
& & \left. + (K^-)^{2} \left[ e^{i(kz-k_g y-\Omega^- t)} -
e^{-i(kz+k_g y+\Omega^- t-2ka)} \right] \right\} \; ,   \label{2.20}
\end{eqnarray}
where
\begin{eqnarray}
& & K^{\pm} \equiv k\pm k_g \; , \label{2.21} \\
& & \Omega^{\pm} \equiv \omega \pm \omega_g \; . \label{2.22}
\end{eqnarray}
The solution of Eqs.~(\ref{2.16})-(\ref{2.19}) and (\ref{2.12}) with
the perfectly reflecting boundary condition at $ z=a$ and to first
order in $ h$ is
\begin{eqnarray}
& & F^{(1)}_{01}=-\, \frac{AE_0}{4kk_g} \left\{ (K^+)^2\left[ e^{i(kz+k_g y-
\Omega^+ t)}-e^{-i(kz-k_g y+\Omega^+ t-2ka)} \right] \right. \nonumber \\
& & \left. -(K^-)^2\left[
e^{i(kz-k_g y-\Omega^- t)}-e^{-i(kz+k_g y+\Omega^- t-2ka)} \right]
\right\} \; , \label{2.23}  \\
& & F^{(1)}_{12}=-\, \frac{AE_0}{4k} \left\{ K^+\left[ e^{i(kz+k_g y-
\Omega^+ t)}-e^{-i(kz-k_g y+\Omega^+ t-2ka)} \right] \right. \nonumber \\
& &  \left. +K^- \left[ e^{i(kz-k_g y-\Omega^- t)}-e^{-i(kz+k_g y+\Omega^-
t-2ka)} \right] \right\} \; , \label{2.24}  \\
& & F^{(1)}_{13}=-\, \frac{AE_0}{4k_g} \left\{ K^+ \left[ e^{i(kz+k_g y-
\Omega^+ t)}+e^{-i(kz-k_g y+\Omega^+ t-2ka)} \right] \right. \nonumber \\
& & \left. -K^- \left[ e^{i(kz-k_g y-\Omega^- t)}
+e^{-i(kz+k_g y+\Omega^- t-2ka)} \right] \right\} \; , \label{2.25}
\end{eqnarray}
and the other $ F^{(1)}_{\mu\nu}=0$.
We can now use this example case to write the solutions to Eqs.~(\ref{2.11})
and (\ref{2.12}) for the actual interferometer arms. \\
\section{Solutions of the Maxwell equations and computation of the phase shift}
{\bf $x$ arm:}\\
By means of the coordinate transformation $ \left\{ x^{\mu} \right\}
\mapsto  \left\{ {x'}^{\mu} \right\} $, with\setcounter{equation}{0}
\begin{eqnarray}
&& ct'=ct \; , \nonumber \\
&& x'=y \; , \nonumber \\
&& y'=z \; , \nonumber \\
&& z'=x \; ,                              \label{2.26}
\end{eqnarray}
the unperturbed electromagnetic field tensor given by
Eqs.~(\ref{2.4}) and (\ref{2.5}) transforms in such a way that its only
nonvanishing components in the $\left\{ x'^{\mu} \right\} $ system are
\be
E'^{(0)}_{x}=-F'^{(0)}_{01}=E_0 \left[ e^{i(kz'-\omega t')}-e^{-i(kz'+\omega
t'-2ka)}\right] \; ,                     \label{2.27}  \ee
\be
H'^{(0)}_{y}=-F'^{(0)}_{13}=E_0 \left[ e^{i(kz'-\omega t')}+e^{-i(kz'+\omega
t'-2ka)}\right] \; .                        \label{2.28}  \ee
The metric perturbation becomes
\be
h'_{11}=-h'_{33}=\frac{A}{2}\left[ e^{i(k_g y'-\omega_g
t')}+e^{-i(k_g y'-\omega_g t')} \right] \; .              \label{2.29}  \ee
Equations (\ref{2.27}), (\ref{2.28}) and (\ref{2.29}) coincide with
Eqs.~(\ref{2.14}), (\ref{2.15}) and (\ref{2.13}), the solutions of
which are given by Eqs.~(\ref{2.23})-(\ref{2.25}). Transforming back
to the $ \left\{ x^{\mu} \right\} $ system, these become
\begin{eqnarray}
& & F^{(1)}_{02}=-E^{(1)}_{y}=-\, \frac{AE_0}{4kk_g} \left\{ (K^+)^2
\left[ e^{i(kx+k_g z-\Omega^+ t)}-e^{-i(kx-k_g z+\Omega^+ t-2ka)} \right]
\right. \nonumber   \\
& & \left. -(K^-)^2\left[ e^{i(kx-k_g z-\Omega^- t)}-
e^{-i(kx+k_g z+\Omega^- t-2ka)} \right] \right\} \; , \label{2.30} \\
& & F^{(1)}_{12}=H^{(1)}_{z}=\frac{AE_0}{4k_g} \left\{ K^+\left[ e^{i(kx+k_g
z-\Omega^+ t)}+e^{-i(kx-k_g z+\Omega^+ t-2ka)} \right] \right.
\nonumber \\
& & \left. -K^- \left[ e^{i(kx-k_g z-\Omega^- t)}+
e^{-i(kx+k_g z+\Omega^- t-2ka)} \right] \right\} \; , \label{2.31} \\
& & F^{(1)}_{23}=H^{(1)}_x=-\, \frac{AE_0}{4k} \left\{ K^+ \left[ e^{i(kx+k_g
z-\Omega^+ t)}-e^{-i(kx-k_g z+\Omega^+ t-2ka)} \right] \right.
\nonumber \\
& & \left. +K^- \left[ e^{i(kx-k_g y-\Omega^- t)}-
e^{-i(kx+k_g z+\Omega^- t-2ka)} \right] \right\} \; ,      \label{2.32}
\end{eqnarray}
and the other $ F^{(1)}_{\mu\nu}=0$.
\\
{\bf $y$ arm:} \\
We now perform the coordinate transformation
\begin{eqnarray}
&& ct'=ct \; , \nonumber \\
&& x'=x \; , \nonumber \\
&& y'=z \; , \nonumber \\
&& z'=-y \; ,                         \label{2.33}
\end{eqnarray}
which gives
\be
E'^{(0)}_{x}=E_0 \, e^{2ika} \left[ e^{i(kz'-\omega t')}-e^{-i(kz'+\omega
t'-2ka')}\right] \; ,                      \label{2.34}  \ee
\be
H'^{(0)}_{y}=E_0 \, e^{2ika}\left[ e^{i(kz'-\omega t')}+e^{-i(kz'+\omega
t'-2ka')}\right] \; ,                   \label{2.35}  \ee
where $ a'\equiv -a $, and the boundary condition $ E^{(0)}_{x}(y=a)=0
$ is transformed to $ E'^{(0)}_{x}(z'=a')=0 $. In addition,
\be
h'_{11}=-h'_{33}=-\, \frac{A}{2} \left[ e^{i(k_g y'-\omega_g t')}+
e^{-i(k_g y'-\omega_g t')} \right] \; .         \label{2.36}  \ee
Thus, we reproduce the case described by Eqs.~(\ref{2.13})-(\ref{2.15}),
provided we make the following substitutions:
\begin{eqnarray}
& & E_0 \mapsto E_0 \, e^{2ika} \; ,          \label{2.37} \\
& & A \mapsto -A \; .                         \label{2.38}
\end{eqnarray}
The electromagnetic field perturbations are then given by
Eqs.~(\ref{2.23})-(\ref{2.25}), with these substitutions. Transforming
back to the $ \left\{ x^{\mu} \right\} $ coordinate system, we find
that the only nonvanishing $ F^{(1)}_{\mu\nu}$ are
\begin{eqnarray}
& & F^{(1)}_{01}=-E^{(1)}_{x}=-\, \frac{AE_0}{4kk_g} \left\{ (K^+)^2
\left[ e^{i(ky+k_g z-\Omega^+ t)}-e^{-i(ky-k_g z+\Omega^+ t-2ka)} \right]
\right. \nonumber \\
& & \left. -(K^-)^2\left[ e^{i(ky-k_g z-\Omega^- t)}-
e^{-i(ky+k_g z+\Omega^- t-2ka)} \right] \right\} \; , \label{2.39} \\
& & F^{(1)}_{12}=H^{(1)}_{z}=-\, \frac{AE_0}{4k_g} \left\{ K^+\left[
e^{i(ky+k_g z-\Omega^+ t)}+e^{-i(ky-k_g z+\Omega^+ t-2ka)} \right]
\right. \nonumber \\
& & \left. -K^- \left[ e^{i(ky-k_g z-\Omega^- t)}+
e^{-i(ky+k_g z+\Omega^- t-2ka)} \right] \right\} \; , \label{2.40} \\
& & F^{(1)}_{13}=-H^{(1)}_y=-\, \frac{AE_0}{4k} \left\{ K^+ \left[ e^{i(ky+k_g
z-\Omega^+ t)}-e^{-i(ky-k_g z+\Omega^+ t-2ka)} \right] \right.
\nonumber \\
& & \left. +K^- \left[ e^{i(ky-k_g z-\Omega^- t)}-
e^{-i(ky+k_g z+\Omega^- t-2ka)} \right] \right\} \; . \label{2.41}
\end{eqnarray}
In the TT gauge, the effect of the gravitational wave is to generate
two sideband components of the electromagnetic signal at frequencies $
\Omega^{\pm} $, which propagate together with the carrier of frequency
$ \omega $. This can be seen as a phase shift at the photodiode, where
the two signals reflected by the mirrors are compared.

In the $x$ arm, in the limit $ \omega >>\omega_g $, the backward propagating
component of the electric field on the $z=0$ plane is
\begin{eqnarray}
E_{yb}=-E_0 \cos(kx+\omega t-2ka)-\frac{AE_0}{4} \frac{k}{k_g} \left[
\cos( kx+\Omega^+ t-2ka) \right. \nonumber \\
\left. -\cos( kx+\Omega^- t-2ka) \right] =-E_0 \cos(kx+\omega t-2ka)
\nonumber \\
+\frac{AE_0}{2} \frac{k}{k_g}\sin (kx+\omega t-2ka) \sin ( \omega_g t) \; .
\label{3.1}
\end{eqnarray}
By setting
\begin{eqnarray}
E_{yb} & \equiv & -E_0 \cos ( kx+\omega t-2ka+\delta\phi_x ) \nonumber \\
& \simeq & -E_0 \cos ( kx+\omega t-2ka)+E_0 \, \delta\phi_x
\sin( kx+\omega t-2ka) \; ,                             \label{3.2}
\end{eqnarray}
and comparing Eqs.~(\ref{3.1}) and (\ref{3.2}) at the arrival time
$\tau$ of the signals at the photodiode ($\tau$ is the round trip time
for the laser light in each of the interferometer arms), we find the
phase shift for the $x$ arm:
\be
\delta \phi_x=\frac{A}{2} \, \frac{\omega}{\omega_g} \sin ( \omega_g
\tau) \; .                          \label{3.3}  \ee
Analogously, we obtain the phase shift for the $y$ arm
\be
\delta \phi_y=-\, \frac{A}{2} \, \frac{\omega}{\omega_g} \sin ( \omega_g
\tau) \; ,                      \label{3.4}   \ee
which is equal and opposite to $\delta \phi_x$. The differential phase shift
between the signals coming from the two interferometer arms is thus
\be
\delta \phi \equiv \delta \phi_x-\delta \phi_y=2A \,
\frac{\omega}{\omega_g} \sin \left( \frac{\omega_g \tau}{2}\right)
\cos\left( \frac{\omega_g \tau}{2} \right) \; ,      \label{3.5} \ee
expressed in a form which is convenient for the purposes of
comparison.
\section{Beam deflection and polarization shift}
We now consider the deflection of the laser beams induced by their
interaction with the gravitational waves. The fact that
gravitational waves curve light rays propagating through them is well
known, and has been studied in the geometric optics approximation by
various authors, mainly in view of its possible astrophysical
consequences \cite{Zipoy}-\cite{Dautcourtb}. To the best of our knowledge,
it has not been discussed previously in the context of interferometry.
In principle, it could be of significance because it implies a
change in the distance of beam traversal and as a consequence, a variation
in phase of the waves. However, we will demonstrate that this phase
shift is negligible in comparison with that discussed in Sec.~3 under
the normal conditions currently considered.

We now analyze the configuration described by Eqs.~(\ref{2.13})-(\ref{2.15}):
The unperturbed Poynting vector of the electromagnetic waves
becomes\setcounter{equation}{0}
\be
\vec{S}^{(0)}=\frac{c}{4\pi}\, \vec{E}^{(0)} \times \vec{H}^{(0)}=
\frac{c}{4\pi} \, E^{(0)}_x H^{(0)}_y \, \vec{e}_z       \label{4.1} \ee
(where $ \vec{e}_i $ denotes the unit vector parallel to the $i$-th
axis). In the presence of the gravitational wave, the Poynting vector
is
\be   \label{4.2}
\vec{S}^{(tot)}=\frac{c}{4\pi}\, \vec{E} \times
\vec{H}=\vec{S}^{(0)}+\vec{S}^{(1)} \; ,   \ee
where
\be         \label{4.3}
\vec{S}^{(1)}=\frac{c}{4\pi} \left[ \left( E^{(0)}_x H^{(1)}_y+E^{(1)}_x
H^{(0)}_y \right) \vec{e}_z-E^{(0)}_xH^{(1)}_z \vec{e}_y \right] +O(2) . \ee
Clearly, the vector $\vec{S}^{(tot)}$ is rotated in
the $(y,z)$ plane with respect to $ \vec{S}^{(0)}$. We consider, for
the sake of simplicity, only the forward propagating component of the
electromagnetic wave. Let $ \theta $ be the angle between the vectors $
\vec{S}^{(0)}_{forward} $ and $ \vec{S}^{(tot)}_{forward} $ in the
$(y,z)$ plane where they lie, i.e. the deflection angle. We have
\be
S^{(tot)}_{y\; forward} =\left| \vec{S}^{(tot)}_{forward} \right|
\sin \theta \; ,  \label{4.4}   \ee
from which we obtain (we omit the subscript ``forward'' in the
following)
\be
\sin \theta =-\,\frac{H^{(1)}_z}{|H^{(0)}_y|} \,\mbox{sign}( E^{(0)}_x)+O(2)
\;.                        \label{4.5}       \ee
In the limit $ \omega>>\omega_g$, in the plane $y=0$, and
expanding $\sin \theta$ to first order, we find
\be   \label{4.6}
\theta =\frac{A}{2} \cos( \omega_g t) \; .               \ee
Note that $ \theta=O(h)$. Equation~(\ref{4.6}) can be compared with the
deflection angle obtained in previous analyses performed in the geometric
optics approximation. If we consider a photon whose unperturbed path is
parallel to the $z$ axis, and with four-momentum
\be                \label{4.7}
p^{\mu}=p^{(0) \mu}+\delta p^{\mu}=(1,0,0,1)+\delta p^{\mu} \; ,  \ee
where $ \delta p^{\mu} $ are small deflections, the equation of null
geodesics give
\be           \label{4.8}
\delta p^i=-\, \frac{1}{2} \int_{S}^{O} dz\, \left(
h_{00}+2h_{03}+h_{33} \right)^{,i}+O(2)         \ee
$(i=1,2$), where the integral is computed along the unperturbed photon
path from the light source $S$ to the observer $O$. This formula can
be found, e.g., in Refs.~\cite{Linder86}-\cite{BertottiCatenacci}.
In our case we obtain, from Eq.~(\ref{2.13}),
\begin{eqnarray}
& & \delta p^x=O(2)                                      \label{4.9} \\
& & \delta p^y=\frac{A}{2} \cos( \omega_g t)+O(2) \; , \label{4.10}
\end{eqnarray}
which confirm that the deflection takes place in the $(y,z)$ plane,
and the value of $\theta$ is as in Eq.~(\ref{4.6}). This provides a check of
our results from an independent source. It is of interest to consider
the additional phase shift at the photodiode due to the change in path
length travelled by the photons under the deflection. If $ \delta l_i$
is the extra length for the
$i$-th arm ($i=1,2$), the phase shift it induces is $\delta
\phi_i=2\pi\delta l_i/\lambda$. It is immediate to see that, if $l_i$
is the total length travelled by the photon,
\be
\delta l_i=l_i-a=l_i(1-\cos \theta ) \simeq a\, \frac{\theta^2}{2} \;
.     \label{4.11}     \ee
The additional phase shift introduced by the deflection of light is
thus $ \delta \phi_i =O(h^2)$, and is completely negligible. This is
true also for the differential phase shift in the two arms of the
interferometer.

An interesting property of the interaction between gravitational
and electromagnetic waves can be seen from the solutions
(\ref{2.23})-(\ref{2.25}) of the coupled Einstein-Maxwell equations. We
note that the unperturbed electric and magnetic fields are polarized along
the $x$
and $y$ axes, respectively. The perturbation to the electric field is
parallel to the $x$ axis, while that of the magnetic field has
components along both the $y$ and $z$ axis. This is consistent with
the result \cite{Faraoni92} that a weak gravitational wave induces no
rotation in the plane of polarization of electromagnetic radiation, to
first order in $h$. In fact, let us consider a given point O on the
light beam, at a given time, and the two planes ${\cal P}_1$ and
${\cal P}_2$ passing through O, which are defined by the pairs of
vectors $(\vec{E}^{(0)},\vec{H}^{(0)})$ and
$(\vec{E}^{(tot)},\vec{H}^{(tot)})$, respectively (see Fig.~2). The
normals to these planes are parallel to the
Poynting vectors $\vec{S}^{(0)}$ and $\vec{S}^{(tot)}$ respectively.
${\cal P}_1$ is associated with the forward propagating electromagnetic
wave at O at the given time, in the unperturbed case (no gravitational
waves) and ${\cal P}_2$ with that corresponding to the perturbed wave.
${\cal P}_1$
and ${\cal P}_2$ intersect along their common $x$ axis. The plane ${\cal P}_2$
can be obtained by rotating ${\cal P}_1$ by an angle $\theta$ around the
$x$ axis. The projection of the magnetic field $H^{(tot)} \in {\cal P}_2$
on ${\cal P}_1$ lies on the direction of $H^{(0)}$ (i.e. on the $y$
axis of ${\cal P}_1$). In this sense, the variation of the direction of
the magnetic field due to the gravitational wave arises solely from
the deflection of the light beam. No rotation of the electric and
magnetic field, with respect to the unperturbed case, takes place in a
plane orthogonal to the Poynting vector (in the sense of the background
Minkowski metric). Although these considerations could be made more
rigorous by using advanced concepts from differential geometry, it is
unnecessary, since a general proof of the fact that
gravitational waves do not rotate the polarization vector of
electromagnetic waves propagating through them, to first order, can be
found in Ref.~\cite{Faraoni92}.
\section{Discussion and conclusion}
In Sec.~3 we have presented a complete solution (to first order in $
h$) to the Maxwell equations
in the field of a weak gravitational wave, taking into account the
boundary conditions. Our result in Eq.~(\ref{3.5}) disagrees with that by
Lobo by the presence of the factor $ \cos ( \omega_g \tau/2)$. This
factor is absent in Lobo's paper \cite{Lobo}, but it does appear in a
paper by Meers \cite{Meers}. This is contrary to Lobo's
claimed agreement with Meers. To reproduce Lobo's result in our
calculations, it would require a time interval $(0,\tau/2 )$ to be
considered twice instead of correctly considering the two distinct
time intervals $( 0, \tau/2 )$ and $( \tau/2, \tau )$. The former
approach would give a factor $2 \sin( \omega_g \tau /2 ) $ in our
Eqs.~(\ref{3.3}) and (\ref{3.4}), and corresponds to computing the
phase shift for the forward propagating signal during the time
interval $(0, \tau/2 )$, and incorrectly doubling the result in order
to obtain the phase shift for the reflected signal travelling from the
mirror to the photodiode during the time interval $( \tau/2, \tau )$.
Apparently, the final formulas for the phase shift in Ref.~\cite{Lobo}
are obtained in this way. The correct computation gives a factor $
\sin( \omega_g \tau )=2 \sin( \omega_g \tau /2) \cos( \omega_g \tau
/2) $ instead of $ 2 \sin( \omega_g \tau /2) $ in Eqs.~(\ref{3.3}) and
(\ref{3.4}). However, it should
be noted that the distinction is irrelevant in the limit $
\tau/T_g <<1 $, wherein both Eq.~(\ref{3.5}) and the result of
Ref.~\cite{Lobo} agree with Ref.~\cite{Thorne}. This limit is not
unphysical: if $ \tau/T_g \sim 1 $, Eq.~(\ref{3.5}) gives $ \delta
\phi \sim 0 $, and the detector is rendered useless due to the fact that
the metric perturbation reverses its sign and the phase shift
accumulated in the first half period of the gravitational wave is
destroyed during the second half \cite{Thorne}. However, the inherent
broadband nature of the detector makes it sensitive to gravitational wave
pulses which are not monochromatic, but have Fourier components in a band
of frequencies $ \Delta \nu_g $, some of which may have $ \omega_g \tau$
less than $1$, but not very small. In such cases the full Eq.~(\ref{3.5})
applies.

Perhaps, the most promising sources of gravitational waves
detectable in the near future with beam detectors are coalescing
binaries containing compact objects. These systems give rise to a
characteristic signal (``chirp'') with frequency increasing up to 1~KHz
in the final stages of coalescence. The final moments of the process
are not completely understood \cite{Nakamura,OoharaNakamura,Kochanek}, and
might well give rise to higher frequencies not satisfying $ \omega_g \tau <<1$.

In Sec.~4, we used the formalism in Ref.~\cite{Cooperstock68} to
deduce the deflection of the laser beams in their interaction with the
gravitational waves. We found that, while this is a first order effect
in $h$, it only induces a second order contribution to the phase
shift.

Finally, we note that if perfectly reflecting mirrors are placed at $x=0$,
$y=0$ normal to the $x$ and $y$ axes, the wave number of the unperturbed
electromagnetic field in the cavity so constructed is restricted to the
values $ k=n\pi /a $, where $n$ is an integer. This system can provide an
idealized description of a laser interferometric detector operating with
Fabry-Perot cavities in its arms (\cite{Thorne}-\cite{Giazotto} and
references therein).
\section*{Acknowledgments}
V. F. acknowledges financial support from the Fondazione Angelo della
Riccia and the warm hospitality at the University of Victoria. This
research was supported, in part, by a grant from the Natural Sciences
and Engineering Research Council of Canada.
\vspace*{1truecm}
{\bf Figure captions:} \\ \\
{\bf Figure~1:} The gravitational wave propagates
along the positive $z$ axis; electromagnetic waves starting from the
beam splitter at the origin are reflected by mirrors placed at $x=a$,
$y=a$. \\ \\
{\bf Figure~2:} The effects induced by the gravitational wave on the plane
of polarization of a forward propagating electromagnetic wave.
$\vec{E}^{(0)}$, $\vec{H}^{(0)}$ and $\vec{S}^{(0)} $ are the electric
and magnetic fields, and the Poynting vector, respectively, in the
unperturbed case. $\vec{E}^{(tot)}$, $\vec{H}^{(tot)}$ and
$\vec{S}^{(tot)} $ are the corresponding quantities in the presence of
gravitational waves. $\theta$ is the deflection angle of the beam. The
planes ${\cal P}_1$ and ${\cal P}_2$ are the polarization planes of the
electromagnetic radiation in the unperturbed and perturbed case.
\end{document}